\documentclass[aps,preprint,amsmath,amssymb]{revtex4}
\usepackage{graphicx}
\def\e{\epsilon}
\def\hmp{\hat m_{P}}
\def\hml{\hat m_{\ell}}

\def\hmv{\hat m_{V}}
\def\ml{m_{\ell}}
\def\hmb{\hat m_{b}}
\def\vp{\tilde{V}_{ub}}
\def\ve{\varepsilon}
\def\msg{M_{\tilde{g}}}
\def\msd{M^2_{\tilde{d}_{L}}}
\def\msu{M^2_{\tilde{u}_{L}}}
\def\msdr{M^2_{\tilde{d}_{R}}}

\begin{document}

\title{\Large \bf Charged Higgs on  $B^{-} \to \tau \bar\nu_{\tau}$
and $\bar B\to  P( V) \ell \bar\nu_{\ell}$}

\date{\today}
\author{ \bf \large Chuan-Hung Chen$^{1,2}$\footnote{Email:
physchen@mail.ncku.edu.tw} and Chao-Qiang
Geng$^{3,4}$\footnote{Email: geng@phys.nthu.edu.tw}
 }

\affiliation{ $^{1}$Department of Physics, National Cheng-Kung
University, Tainan 701, Taiwan \\
$^{2}$National Center for Theoretical Sciences, Taiwan\\
$^{3}$Department of Physics, National Tsing-Hua University, Hsin-Chu
300, Taiwan  \\
$^{4}$Theory Group, TRIUMF, 4004 Wesbrook Mall, Vancouver, V6T 2A3, Canada
 }
\date{\today}
\begin{abstract}
We study the charged Higgs effects on the decays of $B^{-} \to \tau
\bar\nu_{\tau}$ and $\bar B\to  P( V) \ell \bar\nu_{\ell}$ with
$P=\pi^{+},D^{+}$ and $V=\rho^{+},D^{*+}$. We concentrate on the
minimal supersymmetric standard model  with nonholomorphic terms at
a large $\tan\beta$. To extract new physics contributions, we define
several physical quantities related to the decay rate and angular
distributions to reduce uncertainties from the QCD as well as the
CKM elements. With the constraints from the recent measurement on
the decay branching ratio of $B^{-} \to \tau \bar\nu_{\tau}$, we
find that the charged Higgs effects could be large and measurable.

\end{abstract}

\maketitle %
\section{Introduction}

Many exclusive hadronic $B$ decay modes have been
observed in branching ratios (BRs) and CP asymmetries (CPAs) at $B$
factories \cite{HFAG}.
However, it is hard to give conclusive
theoretical predictions for most of the processes
in the standard model (SM)
due to the nonperturbative QCD effects.
Consequently, it is not
easy to tell whether there are some derivations between theoretical predictions and experimental measurements.
To search for new physics,
it is important to look for some observables which contain less
 theoretical uncertainties.
With enormous
$B$ events,
recently, the BELLE \cite{belle} and BABAR \cite{babar} Collaborations
have measured the purely leptonic decay of $B^{-}\to
\tau \bar \nu_{\tau}$ as \cite{Barlow}
\begin{eqnarray}
BR(B^- \to \tau \bar \nu_{\tau})&=&
\left(1.79^{+0.56+0.39}_{-0.49-0.46}\right)\times 10^{-4}\ (BELLE)
\nonumber\\
&=& \left(0.88^{+0.68}_{-0.67}\pm
0.11\right)\times 10^{-4}< 1.8 \times 10^{-4}\,\,  (90\%\,C.L.)\ (BABAR)
\nonumber\\
&=& \left(1.36\pm 0.48\right)\times
10^{-4}\ (BELLE\,+\,BABAR)
\end{eqnarray}
This observation provides a possibility
to detect new physics. It is well known that the SM
contribution to the decay branching ratio arises from the charged weak
interactions
with the main uncertainty from the
Cabibbo-Kobayashi-Maskawa (CKM) matrix element $V_{ub}$
and the B-meson decay constant $f_{B}$.
The value of $V_{ub}$
has been constrained by the inclusive and exclusive charmless
semileptonic $B$ decays,
given by
$|V_{ub}|=(4.39\pm 0.33)\times 10^{-3}$ \cite{HFAG}
and
$|V_{ub}|=(3.67\pm 0.47)\times 10^{-3}$ \cite{PDG04}, respectively.
Obviously,
when $|V_{ub}|$ is fixed, the decay of
$B^{-}\to \tau \bar \nu_{\tau}$ could be used to determine
$f_{B}$, which should not be far away from that calculated
by the lattice QCD \cite{LQCD} as well as
extracted from other
experimental data, such as $\Delta M_{B}$
\cite{PDG04,UTfit}.
 Clearly, if there appears some significant derivation,
 it could imply the existence of  physics beyond the SM.

 The most interesting new physics contribution to the decay is
 the charged Higgs effect at tree level \cite{Hou,AR}.
 Similar effect has also been studied in the inclusive
 \cite{Inclu} and exclusive \cite{Exclu} semileptonic $B$ decays.
  It is known that the charged Higgs boson
 exists in any model with two or more Higgs doublets, such as
 the minimal supersymmetric standard model (MSSM)  which
 contains two Higgs doublets $H_{d}$ and $H_{u}$ coupling to
  down and up type quarks, respectively.
  In the MSSM, it is natural to avoid the flavor changing neutral
current (FCNC) at tree level. However, due to supersymmetric
breaking effects, it is found that in the large
$\tan\beta$ region, the contribution to
the down type quark masses from
the nonholomorphic terms $Q D^c
H_{u}$ generated at one-loop could be as large as that
 from the holomorphic ones $Q D_{c} H_{d}$
\cite{SUSY1,SUSY2}. Subsequently, many interesting Higgs related
phenomena have been studied \cite{SUSY3,SUSY4,SUSY5,SUSY6,SUSY7}.

In this
paper, we will study the charged Higgs contributions  with
the nonholomophic corrections to the leptonic decays of $B^{-}\to \ell
\bar\nu_{\ell}$ and the exlusive semileptonic decays of
$B\to P(V) \ell \bar \nu_{\ell}$  where $\ell$ denote as the charged leptons and $P$ ($V$) stand for
the pseudoscalar (vector) mesons.
 In particular, we will investigate
the differential decay rates
and the lepton angular distributions
 in the exclusive semileptonic modes
 to examine the charged Higgs effects based on the constraint
 from the measurement on $BR(B^- \to \tau \bar \nu_{\tau})$.

The paper is organized as follows. In Sec.~\ref{sec:chiggs}, we
derive the couplings of charged Higgs to quarks by including the
one-loop corrections to the Yukawa sector. In Sec. III, we present the
formalisms for the decay rates of $B^{-}\to \ell \bar \nu_{\ell}$,
the differential decay rates and angular asymmetries of $B\to
P(V)\ell \bar \nu_{\ell}$ in the presence of the charged Higgs contributions.
We display the numerical analysis in
Sec.~\ref{sec:numerical}. Finally, we summarize the results in
Sec.~\ref{sec:conclusion}.

\section{Couplings of Charged Higgs to quarks} \label{sec:chiggs}

In models with two Higgs doublets,  the general
 Yukawa couplings with radiative corrections
 for the quark sector under
  the gauge groups $SU(2)_L\times U(1)_{Y}$
can be written as \cite{SUSY6}
\begin{eqnarray}
-{\cal L}_{Y}=\bar Q_L \left[{\bf H_{d}}+ \left( \epsilon_{0}
+\epsilon_{Y} Y_{u} Y^{\dagger}_{u}\right){\bf \tilde H_{u}}
\right]  {\bf Y_{d}} D_{R}  +h.c. \label{eq:yukawa}
\end{eqnarray}
where $Q^{T}_{L}=(U,D)_{L}$ and $D_{R}$ denote the $SU(2)$ doublet
and singlet of quarks, respectively, ${\bf H^{T}_{d}}=(\phi^{+}_{d},
\phi^{0}_{d})$ and ${\bf \tilde H_{u}}=-i\tau_{2} {\bf H^*_{u}}$
with ${\bf H^{T}_{u}}=(\phi^{*0}_{u}, -\phi^{-}_{u})$ are the two
Higgs doublets, ${\bf Y_{d(u)}}$ is the $3\times 3$ Yukawa mass
matrix for down (up) type quarks, and $\e_{0,Y}$ stand for the
effects of radiative corrections. Since only the down-type quark
mass matrix can have large radiative corrections, we will not
address the parts related to $\bar Q_{L} U_{R}$. Moreover, for
simplicity, we choose
 ${\bf Y_{d}}$ to be a diagonal matrix
${\bf Y_{d}}_{ij}= y_{di}\delta_{ij}$ while ${\bf Y_{u}}$ is
diagonalized by $V^{0L}_{U} {\bf Y_{u}} V^{0R\dagger}_{U}\equiv
U=diag\{y_{u},y_{c},y_{t}\}$ with $V^{0L(R)}_{U}$ being unitary matrices.
In terms of the charged weak interaction, denoted by $I_{W}=\bar U_L
\gamma^{\mu} D_{L} W^{+}_{\mu}$,
the Cabibbo-Kobayashi-Maskawa (CKM) matrix is $V^{0}= V^{0L}_{U}$.
 From Eq.~(\ref{eq:yukawa}), we know that due to the
appearance of $\e_{Y}$, the down-type quark mass matrix, expressed by
\begin{eqnarray}
M_{D}&=& \left[ 1+\tan\beta \left(
\e_{0}+\e_{Y}V^{0\dagger}UU^{\dagger}V^{0}
\right)\right]{\bf Y_{d}} {\rm v_d}\, ,\nonumber \\
&=& M^{\rm dia}_{D} + \delta M_{D}\,, \label{eq:mass_d}
\end{eqnarray}
is no longer diagonal, where
\begin{eqnarray}
M^{\rm dia}_{Di}&=& y_{di} {\rm v_d} [1+\tan\beta \e_i]\, , \nonumber \\
\delta M_{Dij}&=& y_{dj} {\rm v_d} \tan\beta \epsilon_{Y}
y^{2}_{t} V^{0}_{tij}\,,
\end{eqnarray}
with ${\rm v_{d(u)}}=\langle \phi^{0}_{d(u)} \rangle$,
$\tan\beta=v_{u}/v_{d}$, $V^{0}_{tij}=V^{0*}_{ti} V^{0}_{tj}$ and
$\e_i=\epsilon_{0}+ \epsilon_Y y^{2}_{t} \delta_{i3}$. Here we have
neglected the contributions of $y_{u(c)}$ due to the hierarchy
$y_{u}\ll y_{c}\ll y_{t}$.

In order to diagonalize the mass matrix of Eq.~(\ref{eq:mass_d}),
we need to introduce new unitary matrices $V^{L(R)}_{D}$ so that
the physical states are given by
\begin{eqnarray}
d_{L}= V^L_{D} D_{L}, \ \ \ d_{R}=V^{R}_{D} D_{R},
\label{eq:unitary}
\end{eqnarray}
and the diagonalized mass matrix is $m_{D}=V^{L}_{D} M_{D} V^{R\dagger}_{D}$.
Subsequently, we have the relationships
\begin{eqnarray}
m_{D} m^{\dagger}_{D} &=& V^{L}_{D} M_{D} M^{\dagger}_{D}
V^{L\dagger}_{D}\, , \nonumber \\
m^{\dagger}_{D} m_{D} &=& V^{R}_{D} M^{\dagger}_{D} M_{D}
V^{R\dagger}_{D}\, . \label{eq:mass2}
\end{eqnarray}
Since the off-diagonal terms in Eq.~(\ref{eq:mass_d}) are
 associated with $\e_{Y}$ which is much
less than unity, we can find $V^{L(R)}_{D}$ by the perturbation
in $\e_{Y}$. At the leading $\e_{Y}$, the unitary
matrices could be expressed by $V^{L}_{D}\approx 1+\Delta^{L}_{D}$
and $V^{R}_{D}\approx 1+\Delta^{R}_{D}$. By Eq.~(\ref{eq:mass2}),
we easily obtain
\begin{eqnarray}
\Delta^{L}_{Dij[i\neq j]}= \frac{M^{\rm dia}_{Di} (\delta
M^{\dagger}_{D})_{ij} + \delta M_{Dij} M^{\rm dia}_{Dj}}{|M^{\rm
dia}_{Di}|^2 - |M^{\rm
dia}_{Dj}|^2}\, , \nonumber \\
\Delta^{R}_{Dij[i\neq j]}= \frac{M^{\rm dia}_{Di} \delta M_{Dij} +
(\delta M^{\dagger}_{D})_{ij} M^{\rm dia}_{Dj}}{|M^{\rm dia}_{Di}|^2
- |M^{\rm dia}_{Dj}|^2}\, .
\end{eqnarray}
We note that $m_{D} m^{\dagger}_{D} \approx  M^{\rm dia}_{D} M^{\rm
dia \dagger}_{D}$.

After getting the unitary matrices $V^{L(R)}_{D}$,  we now
discuss the charged Higgs couplings. According to
Eq.~(\ref{eq:yukawa}), the Yukawa couplings for the charged scalars
 are written as
\begin{eqnarray}
-{\cal L}^{H^+}_{Y}&=& \bar u_{L} V^{0} {\bf Y_{d}} D_{R}
\phi^{+}_{d} + \bar u_{L} V^{0} \left(\epsilon_{0} +
\e_{Y}V^{0\dagger} UU^{+} V^{0} \right){\bf Y_{d}} D_{R}
\phi^{+}_{u}\,.
\end{eqnarray}
In terms of Eq.~(\ref{eq:unitary}), the charged
scalar interactions become
\begin{eqnarray}
-{\cal L}^{H^+}_{Y}&=& \bar u_{L} V^{0} {\bf Y_{d}}V^{R\dagger}_{D} d_{R}
\left(\phi^{+}_{d}-\frac{1}{\tan\beta} \phi^{+}_{u}\right)
+ \frac{1}{{\rm v_d} \tan\beta}\bar u_{L}  V {\bf m_{D}} d_{R} \phi^{+}_{u}. \label{eq:charged_1}
\end{eqnarray}
With the new physical states, the CKM matrix is modified to be
$V=V^{0} V^{L\dagger}_{D}$. Consequently, the first term of
Eq.~(\ref{eq:charged_1}) could be expressed by the corrected CKM matrix
as $ V^{0} {\bf Y_{d}}V^{R\dagger}_{D}=  V V^{L}_{D} {\bf
Y_{d}}V^{R\dagger}_{D}$. Taking the leading effects of $\e_{Y}$, we
 get
\begin{eqnarray}
 V^{L}_{D} {\bf Y_{d}}V^{R\dagger}_{D}&=&  \left(1+\Delta^{L}_{D}\right) {\bf Y_{d}}
\left( 1-\Delta^{R}_{D}\right)\approx  {\bf Y_{d}} +  \Delta^{L}_{D} {\bf Y_{d}}
- {\bf Y_{d}} \Delta^{R}_{D} \label{eq:charged_2}
\end{eqnarray}
where
\begin{eqnarray}
\left( \Delta^{L}_{D} {\bf Y_{d}}
- {\bf Y_{d}} \Delta^{R}_{D} \right)_{ij[i\neq j]}=-\frac{\e_Y \tan\beta y^{2}_t}
{{\rm v_{d}} \left(1+\tan\beta \e_3 \right) \left( 1+\tan\beta \e_0 \right)} V^{0\dagger}_{i3}V^{0}_{3j}\, .
 \label{eq:charged_3}
\end{eqnarray}
Since Eq.~(\ref{eq:charged_3}) depends on the CKM matrix elements at the lowest order, by
$V=V^{0} V^{L\dagger}_{D}\approx V^{0}(1-\Delta^{L}_{D})$, we obtain the relation to the corrected
CKM matrix elements as
\begin{eqnarray}
V^{0\dagger}_{i3}V^{0}_{33}= V^{\dagger}_{i3} V_{33} \frac{1+\tan\beta \e_3}{1+\tan\beta \e_0}\, .
\label{eq:charged_4}
\end{eqnarray}
It is known that the charged Goldstone and Higgs bosons are given by
\cite{Higgs}
\begin{eqnarray}
G^{+}&=& \cos\beta \phi^{+}_{d} +\sin\beta \phi^{+}_{u}\, \nonumber \\
H^{+}&=& -\sin\beta \phi^{+}_{d} + \cos\beta \phi^{+}_{u}\, .
\end{eqnarray}
Hence, with Eqs.~(\ref{eq:charged_1})$-$(\ref{eq:charged_4}), the
effective interactions for the charged Higgs coupling to $b$-quark and
$q$ with $q=(c,\, u)$ can be written as
\begin{eqnarray}
{\cal L}^{H^+}_{Y}=\left(2\sqrt{2} G_{F} \right)^{1/2}\tilde{V}_{qb} m_{b}\tan\beta
\bar q_{L} b_{R} H^{+}+h.c.
\label{eq:CH_1}
\end{eqnarray}
with
\begin{eqnarray}
\tilde{V}_{qb}=V_{qb}\left[ \frac{1}{1+\tan\beta \e_3}-\frac{\e_Y
y^2_t}{\sin\beta \cos\beta (1+\tan\beta \e_0)^2}\right]\, .
\label{eq:CH_2}
\end{eqnarray}
It is easy to check that when $\e_0$ and $\e_{Y}$ vanish, the couplings return to the ordinary
results with $\tilde{V}_{qb}=V_{qb}$.

In the MSSM, the one-loop corrections to $\e_{0}$ and $\e_{Y}$ are
given by \cite{SUSY2}
\begin{eqnarray}
\e_{0}&=&\frac{2\alpha_{s}}{3\pi} \frac{\mu \msg}{\msd }F_{2}\left(
\frac{\msg^2}{\msd},\frac{\msdr}{\msd}\right)\, ,\ \ \ \e_{Y} =
\frac{1}{(4\pi)^2} \frac{\mu A_{u}}{\msu} F_{2}\left(
\frac{\msg^2}{\msd},\frac{\msdr}{\msd}\right)
\label{Eq:susy}
\end{eqnarray}
with
\begin{eqnarray*}
F_{2}(x,y)=-\frac{x\ln(x)}{(1-x)(x-y)}-\frac{y\ln(y)}{(y-1)(x-y)}\,
,
\end{eqnarray*}
where $\mu$ is the parameter describing the mixing of $H_{d}$ and
$H_{u}$, $A_{U}$ denotes the soft trilinear coupling and
$M_{\tilde{f}}$ with $f=g,\, u_{L},\, d_{R},\, d_{L}$ represent the
masses of the corresponding sfermions.

\section{formalisms for the decays $B^{-}\to \ell \bar \nu_{\ell}$
and $\bar B\to P(V) \ell \bar \nu_{\ell}$} \label{sec:formalism}

In this section, we study the influence of the charged Higgs
on the leptonic
$B^{-}\to \ell \bar \nu_{\ell}$ decays and semileptonic
$\bar B\to P(V) \ell \bar \nu_{\ell}$  decays,
which are governed by $b\to q \ell \bar \nu_{\ell}$ with $q=(c,\, u)$
at the quark level.
The effective Hamiltonian for $b\to q \ell \bar{\nu}_{\ell}$
 with the charged Higgs contribution is given by
\begin{eqnarray}
H_{\rm eff}&=& \frac{G_FV_{ub}}{\sqrt{2}} \left[ \bar
q\gamma_{\mu} (1-\gamma_5) b\, \bar \ell \gamma^{\mu} (1-\gamma_5)
\nu_{\ell} - \delta_{H} \bar q (1+\gamma_5) b\, \bar \ell
(1-\gamma_5) \nu_{\ell}\right] \label{eq:heff}
\end{eqnarray}
with
\begin{eqnarray}
\delta_{H}=\frac{\vp}{V_{ub}} \frac{m_b
\ml\tan^2\beta}{m^{2}_{H^+}}\, .
\end{eqnarray}
Based on the effective interaction in Eq.~(\ref{eq:heff}), in the
following we discuss the relevant physical quantities for various
$B$ decays.

\subsection{Decay rate for $B^{-}\to \ell \bar \nu_{\ell}$}

In terms of Eq.~(\ref{eq:heff}), the transition amplitude for $B^{-}\to \ell \bar \nu_{\ell}$
is given by
\begin{eqnarray}
\langle \ell \bar \nu_{\ell}| H_{\rm eff}| B^{-}\rangle &=&
\frac{G_{F}}{\sqrt{2}} V_{ub}
\left[  \langle0| \bar u \gamma_{\mu}(1-\gamma_5) b | B^- \rangle
\bar \ell \gamma^{\mu} (1-\gamma_5)\nu_{\ell} \right. \nonumber \\
&& \left. -\delta_{H} \langle0| \bar u (1+\gamma_5) b | B^- \rangle
\bar \ell (1-\gamma_5)\nu_{\ell} \right]\,. \label{eq:lepton}
\end{eqnarray}
Since the process is a leptonic decay, the
QCD effect
is only related to the decay constant of the $B$ meson,
which is associated the axial vector current,
defined by
\begin{eqnarray}
\langle 0 | \bar u \gamma^{\mu}(1-\gamma_5) b | B^{-}\rangle
&=&-if_{B}p^{\mu}_{B}\, . \label{eq:dc1}
\end{eqnarray}
By equation of motion,
one has
\begin{eqnarray}
\langle 0 | \bar u \gamma_5 b | B^{-}\rangle &\approx&
-if_{B}\frac{m^{2}_{B}}{m_{b}}\, \label{eq:dc2}
\end{eqnarray}
for the pseudoscalar current.
 From Eqs.~(\ref{eq:lepton}), (\ref{eq:dc1}) and
(\ref{eq:dc2}), the decay rate for $B^{-}\to \ell \bar \nu_{\ell}$
with the charged Higgs contribution is expressed by
\begin{eqnarray}
{\Gamma^{H^{+}}(B^{-}\to \ell \bar\nu_{\ell}) \over
\Gamma^{SM}(B^{-}\to \ell
\bar\nu_{\ell})}=\left|1-\delta_{H}\frac{m^{2}_{B}}{m_{\ell}m_b}
\right|^2\, \label{eq:tau_new}
\end{eqnarray}
where
\begin{eqnarray}
\Gamma^{SM}(B^{-}\to \ell \bar\nu_{\ell})= \frac{G^{2}_{F}
|V_{ub}|^2}{8\pi} f^2_{B} m^2_{\ell} m_{B}
\left(1-\frac{m^2_{\ell}}{m^2_{B}} \right)^2\, . \label{eq:tau_sm}
\end{eqnarray}

\subsection{Differential decay rate and
angular asymmetry for $\bar B\to P \ell \bar \nu_{\ell}$}

By using the effective interaction for $b\to q \ell \bar \nu_{\ell}$
in Eq. (\ref{eq:heff}), we write the decay amplitude
for $\bar B\to P \ell \bar \nu_{\ell}$ to be
\begin{eqnarray}
M(\bar B\to P \ell \bar \nu_{\ell})&=&\langle  \ell \bar\nu_{\ell} P| H_{\rm eff}|\bar B\rangle =
\frac{G_F V_{qb}}{\sqrt{2}} \left[
 \langle P| \bar q \gamma_{\mu}(1-\gamma_5)b| \bar B\rangle \bar \ell\gamma^{\mu} (1-\gamma_5) \nu_{\ell}
 \right. \nonumber \\
 && \left. -\delta_{H}
 \langle P| \bar q (1+\gamma_5)b| \bar B\rangle \bar \ell (1-\gamma_5) \nu_{\ell}
\right] \,. \label{eq:amp_p}
\end{eqnarray}
To get the hadronic QCD effect,
we parametrize the $\bar B\to P$ transition as
\begin{eqnarray}
\langle P(p_{P}) | \bar q \gamma^{\mu}  b| \bar B(p_B)\rangle
&=&
f^{P}_{+}(q^2)\left(P^{\mu}-\frac{P\cdot q}{q^2}q^{\mu}
\right)+f^{P}_{0}(q^2) \frac{P\cdot q}{q^2} q_{\mu}\,,
 \label{eq:bpff}
\\
\langle P(p_{P}) | \bar q \,  b| \bar B(p_B)\rangle
&\approx& f^{P}_{0}(q^2)
\frac{P\cdot q}{m_{b}}\,,
\end{eqnarray}
with $P=p_{B}+p_{P}$ and $q=p_{B}-p_{P}$.
To calculate the decay rate,
we choose the coordinates for
various particles
as follows:
\begin{eqnarray}
q^{2}&=&(\sqrt{q^2},\,0,\, 0,\, 0), \ \ \ p_{B}=(E_{B},\, 0,\, 0,\, |\vec{p}_{P}|), \nonumber \\
p_{P}&=& (E_{P},\, 0,\, 0,\, |\vec{p}_{P}|), \ \ \
p_{\ell}=(E_{\ell},\, |\vec{p}_{\ell}| \sin\theta,\, 0,\, |\vec{p}_{\ell}| \cos\theta)\,,
\label{eq:coordinates}
\end{eqnarray}
where $E_{P}=(m^{2}_{B}-q^2-m^2_{P})/(2\sqrt{q^2})$,
$|\vec{p}_{P}|=\sqrt{E^2_{P}-m^2_{P}}$,
$E_{\ell}=(q^2+m^2_{\ell})/(2\sqrt{q^2})$ and
$|\vec{p}_{\ell}|=(q^2-m^2_{\ell})/(2\sqrt{q^2})$. It is clear that
$\theta$ is defined as the polar angle of the lepton momentum
relative to the moving direction of the $B$-meson in the $q^2$ rest
frame.
The differential decay rate for $\bar B\to
P \ell \bar\nu_{\ell}$ as a function of $q^2$ and $\theta$ is given by
\begin{eqnarray}
\frac{d\Gamma_P}{dq^2 d\cos\theta}&=& \frac{G^{2}_{F} |V_{ub}|^2
m^3_{B}}{2^8
\pi^3}\sqrt{(1-s+\hmp^2)^2-4\hmp^2}\left(1-\frac{\hml^2}{s}
\right)^2 \nonumber \\
&& \times \left[ \Gamma^{P}_{1}+\Gamma^{P}_{2} \cos\theta +
\Gamma^{P}_{3} \cos^2\theta \right]\,,
\label{eq:diff_P}
\\
\Gamma^{P}_{1}&=& f^{P2}_{+}(q^2) \hat
P^{2}_{P}+\hml^2s\left|\frac{1-s-\hmp^2}{s}f^{P}_{+}(q^2)
+C_{2}\right|^2\, ,\nonumber \\
\Gamma^{P}_{2}&=& 2\hml^2\hat P^{2}_{P}\left[ f^{P}_{+}(q^2)
C_{2}-\frac{1-s-\hmp^2}{s}f^{P2}_{+}(q^2)  \right]\,
,\nonumber \\
\Gamma^{P}_{3}&=& -f^{P2}_{+}(q^2) \hat
P^{2}_{P}\left(1-\frac{\hml^2}{s} \right)\,,
 \label{eq:gamma_P}
\end{eqnarray}
where $s=q^{2}/m_{B}^{2}$, $\hat m_{i}=m_{i}/m_{B}$ and
 \begin{eqnarray}
 \hat P_{P}&=&2\sqrt{s} |\vec{p}_{P}|/m_{B}=\sqrt{(1-s-\hmp^2
)^2-4s\hmp^2}\,,
\nonumber\\
C_{2}&=&f^{P}_{+}(q^2)+\left(f^{P}_0(q^2) - f^{P}_+(q^2) \right)
\frac{1-\hmp^2 }{s}-\delta_{H}\frac{1-\hmp^2}{\hml
\hmb}f^{P}_{0}(q^2)\,
\end{eqnarray}

Since the differential decay rate in Eq.~(\ref{eq:diff_P}) involves
the polar angle of the lepton, we can define an angular asymmetry to
be
\begin{eqnarray}
{\cal A}(q^2)={\int^{\pi/2}_{0}d\cos\theta
d\Gamma/(dq^2d\cos\theta)-\int^{\pi}_{\pi/2}d\cos\theta
d\Gamma/(dq^2d\cos\theta) \over \int^{\pi/2}_{0}d\cos\theta
d\Gamma/(dq^2d\cos\theta)+\int^{\pi}_{\pi/2}d\cos\theta
d\Gamma/(dq^2d\cos\theta)}\, .\label{asy}
\end{eqnarray}
 Explicitly, for $\bar B\to P \ell \bar \nu_{\ell}$,
 the asymmetry
is given by
\begin{eqnarray}
{\cal A}_{P}(s)=-{\Gamma^{P}_{2} \over 2 \Gamma^{P}_{1}+ 2/3
\Gamma^{P}_{3}}\, . \label{asy_P}
\end{eqnarray}

\subsection{Differential decay rate and angular asymmetry for
$\bar B\to V \ell \bar \nu_{\ell}$}

Similar to Eq.~(\ref{eq:amp_p}), for $\bar B\to V
\ell \bar \nu_{\ell}$, we need to know the form factors in the
$B\to V$ transition. As usual, we parametrize the transition form
factors to be
\begin{eqnarray}
\langle V(p_{V},\ve )| \bar q \gamma_{\mu } b| \bar{B}%
(p_{B})\rangle &=&i\frac{V^{V}(q^{2})}{m_{B}+m_{V}}\e_{\mu
\alpha \beta \rho }\ve^{*\alpha }P^{\beta }q^{\rho },  \nonumber \\
\langle V(p_{V},\ve )| \bar q \gamma^{\mu} \gamma_{5} b| \bar{B}
(p_{1})\rangle &=&2m_{V}A^{V}_{0}(q^{2})\frac{\ve ^{*}\cdot q}{%
q^{2}}q_{\mu }+( m_{B}+m_{V}) A^{V}_{1}(q^{2})\Big( \ve
_{\mu }^{*}-\frac{\ve ^{*}\cdot q}{q^{2}}q_{\mu }\Big)  \nonumber \\
&&-A^{V}_{2}(q^{2})\frac{\ve ^{*}\cdot q}{m_{B}+m_{V}}\Big( P_{\mu }-%
\frac{P\cdot q}{q^{2}}q_{\mu }\Big) \, .\label{ffv}
\end{eqnarray}
By equation of motion, we have
\begin{eqnarray}
\langle V(p_{V},\ve) | \bar q \gamma_{5} b | \bar
B(p_{B})\rangle=-\frac{2m_{V}}{m_{b}} \ve^*\cdot q
A^{V}_{0}(q^2)\,.
\end{eqnarray}
Consequently, the decay amplitude
 is expressed by
\begin{eqnarray}
M(\bar B\to V \ell \bar \nu_{\ell}) &=& \frac{G_F V_{ub}}{\sqrt{2}}
\left[ T_{\mu} \bar \ell \gamma^{\mu} (1-\gamma_5) \nu_{\ell} +
2\frac{\ve^*\cdot q}{m_{B}}L_{1} \bar\ell \not p_{V} (1-\gamma_5) \nu_{\ell}
\right. \nonumber \\
&& \left.+ m_{\ell} \frac{\ve^*\cdot q}{m_{B}}L_{2} \bar \ell (1-\gamma_5) \nu_{\ell} \right]
\end{eqnarray}
where
\begin{eqnarray}
T_{\mu}&=& i \frac{2V^{V}(q^2)}{m_{B}+m_{V}}\ve_{\mu \alpha \beta
\rho} \ve^{*\alpha} p^{\beta}_{K} q^{\rho} -
(m_{B}+m_{V})A^{V}_{1}(q^2) \left(\ve^{*}_{\mu}
 -\frac{\ve^*\cdot q}{q^2}q_{\mu} \right), \nonumber \\
L_1&=&\frac{A^{V}_2(q^2)}{1+\hmv}, \ \ \
L_{2}=\frac{1-(1-\hmv^2)/s}{1+\hmv} A^{V}_{2}(q^2)- 2\hmv\left(
\frac{1}{s} -\frac{\delta_{H}}{\hml \hmb} \right)A^{V}_{0}(q^2) \,
.
\end{eqnarray}
The differential decay rate
for $\bar B\to  V \ell
\bar\nu_{\ell}$  as a function of $q^2$ and $\theta$ is given by
\begin{eqnarray}
\frac{d\Gamma_V}{dq^2 d\cos\theta}&=& \frac{G^{2}_{F} |V_{ub}|^2
m^3_{B}}{2^8
\pi^3}\sqrt{(1-s+\hmv^2)^2-4\hmv^2}\left(1-\frac{\hml^2}{s}
\right)^2 \nonumber \\
&& \times \left[ \Gamma^{V}_{1}+\Gamma^{V}_{2} \cos\theta +
\Gamma^{V}_{3}\cos^2\theta+  \Gamma^{V}_{4}\sin^{2}\theta \right]
\label{eq:diff_V}
\end{eqnarray}
where
\begin{eqnarray*}
\Gamma^{V}_{1}&=&  s\left[ 2\left(
\frac{V^{V}(q^2)}{1+\hmv}\right)^2 \hat P^2_{V}
+\left(3+\frac{\hat P^2_{V}}{4s\hmv^2} \right)(1+\hmv)^2
A^{V2}_{1}(q^2)\right]\nonumber \\
&&+s L^2_{1}
\left[\frac{E^2_{V}}{m^2_{V}}-1 \right]\left[ \hat P^2_{V}
\left(1+\frac{\hml^2}{s} \right)+4\hml^2\hmv^2\right]\nonumber \\
&&-2s\left(1-s-\hmv^2\right)\left[\frac{E^2_V}{m^2_{V}}-1\right](1+\hmv)A^{V}_1(q^2)L_{1}\nonumber
\\
 &&+ \hml^2 s^2 \left[\frac{E^2_{V}}{m^2_{V}} -1\right] L^2_2
+2\hml^2 s (1-s-\hmv^2) \left[\frac{E^2_{V}}{m^2_{V}} -1\right]L_1
L_2\, ,
\end{eqnarray*}
\begin{eqnarray*}
\Gamma^{V}_{2}&=&16 \hat P_{V}V^{V}(q^2)A^{V}_{1}(q^2)+
2\hml^2(1-s-\hmv^2)\hat P_{V}\left[ \frac{
E^2_{V}}{m^2_{V}}-1\right]L^2_{1} \nonumber
\\
&+&2\hml^2s\hat P_{V}\left[ \frac{ E^2_{V}}{m^2_{V}}-1\right] L_1
L_2 - \frac{\hml^2}{2s\hmv^2}(1-s-\hmv^2)^2 \hat P_{V} (1+\hmv) A^{V}_{1}(q^2) L_{1} \nonumber \\
 &-&\frac{\hml^2}{2\hmv^2}\left(1-s-\hmv^2\right)\hat
P_{V}(1+\hmv)A^{V}_1(q^2)L_{2}\, ,
\end{eqnarray*}
\begin{eqnarray*}
\Gamma^{V}_{3}&=&-s\left(1-\frac{\hml^2}{s} \right) \left[
\frac{E^{2}_{V}}{M^{2}_{V}}(1+\hmv)^2A^{V2}_{1}(q^2)+  \hat
P^2_{V}\left( \frac{ E^2_{V}}{m^2_{V}}-1 \right) L^2_{1} \right]
\nonumber \\
&&+\frac{\hat
P^2_{V}}{2\hmv^2}\left(1-\frac{\hml^2}{s}\right)(1-s-\hmv^2)(1+\hmv)A_1(q^2)L_{2}
\end{eqnarray*}
\begin{eqnarray*}
\Gamma^{V}_{4}&=& -s\left(1-\frac{\hml^2}{s} \right)\left[\hat
P^2_{V}\left(\frac{V^{V}(q^2)}{1+\hmv}\right)^2 + (1+\hmv)^2
A^{V2}_{1}(q^2) \right]
\end{eqnarray*}
with $\hat P_{V}=2\sqrt{s} |\vec{p}_{V}|/m_{B}=\sqrt{(1-s-\hmv^2
)^2-4s\hmv^2}$.
In addition,
 from Eqs.~(\ref{asy}) and (\ref{eq:diff_V}), we obtain the angular asymmetry
for $\bar B\to V \ell \bar \nu_{\ell}$ to be
\begin{eqnarray}
{\cal A}_{V}(s)=-{\Gamma^{V}_{2} \over 2 \Gamma^{V}_{1}+ 2/3 \left(
\Gamma^{V}_{3} +2 \Gamma^{V}_{4}\right) }\, .
\end{eqnarray}

\section{Numerical analysis} \label{sec:numerical}

In the numerical calculations, the model-independence
inputs are used as follows: $G_{F}=1.166\times 10^{-5}$ GeV$^{-2}$,
$m_{b}=4.4$ GeV, and $m_{B}=5.28$ GeV. In addition,
to reduce the unknown parameters in Eq. (\ref{Eq:susy}) for the MSSM,
 we set $|\mu|
\approx |A_{U}|\equiv \bar{\mu}$ and $M_{\tilde{d}_{L}}\approx M_{\tilde{d}_{R}}
\approx M_{\tilde{u}_{L}}\approx M_{\tilde{g}}=M_{S}$ so that the
loop integral is simplified to be a constant with $F(x,y)=1/2$.
Subsequently, the one-loop corrected effects are simplified as
\begin{eqnarray}
\e_{0}&\approx& \pm \frac{\alpha_{s}}{3\pi} \frac{\bar{\mu}}{M_{S} }\, ,\
\ \ \e_{Y} \approx \pm \frac{1}{2(4\pi)^2} \frac{\bar{\mu}^2}{M^2_S}\,,
\label{eq:loopeff}
\end{eqnarray}
where the signs depend on $\mu$ and $A_{U}$, respectively. Hence, we have four possibilities for the sign
combinations in $\e_0$ and $\e_{Y}$. Clearly, based on the
assumption, besides $\tan\beta$ and the charged Higgs mass, now only
one new parameter, denoted by $X=\bar{\mu}/M_{S}$, is introduced in the
charged Higgs couplings. To study the charged Higgs effects at a
large $\tan\beta$ region, we fix $\tan\beta=50$.

\subsection{ $B^{-}\to \tau \nu_{\tau}$}

According to Eq.~(\ref{eq:tau_sm}), it is clear that the BR for
$B^{-}\to \tau \nu_{\tau}$ in the SM depends on two main parameters
$f_{B}$ and $V_{ub}$. To see their contributions, we
calculate the BR with different values of $f_{B}$. For
each value of $f_{B}$, we consider two sets of
$V_{ub}$,
$i.e.$, $(4.39\pm 0.33)\times 10^{-3}$ \cite{HFAG} and
$(3.67\pm 0.47) \times 10^{-3}$ \cite{PDG04},
extracted from the inclusive and exclusive semileptonic $B$ decays,
respectively.

\begin{figure}[htbp]
\includegraphics*[width=4. in]{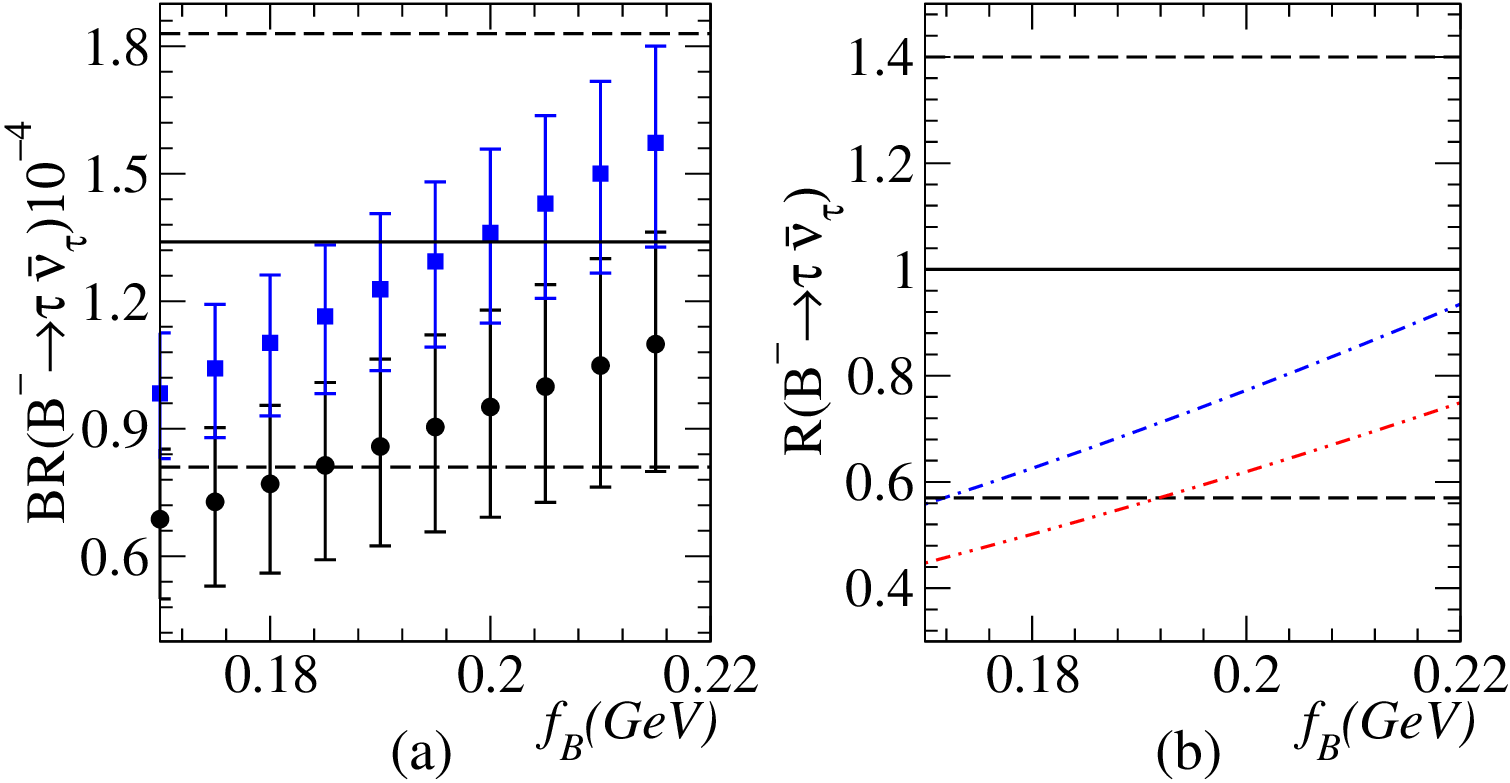}
\caption{(a) BR (in units of $10^{-4}$) and (b) $R$ calculated by
LFQM (dot-dashed) and LCSRs (dot-dot-dashed) for the decay of
$B^{-}\to \tau \bar\nu_{\tau}$ with respect to $f_{B}$, where the
squares and circles stand for $|V_{ub}|=(4.39\pm 0.33)\times
10^{-3}$ and $(3.67\pm 0.47)\times 10^{-3}$ and the solid and dashed
lines represent the central value and
 the $1\sigma$ errors of
the data, respectively.
 }
 \label{fig:error}
\end{figure}

We present the results in Fig.~\ref{fig:error}(a) where the squares
(circles) in the central values denote those calculated with the
bigger (smaller) value of $V_{ub}$ and the solid line displays the
central value of the data, while the dashed lines are the upper and
lower values with $1\sigma$ errors, respectively.
 From the figure, we notice that with the smaller $V_{ub}$, the value
of $f_{B}=0.216 \pm 0.022$ GeV given by the unquenched lattice is
still favorable \cite{HPQCD}. To reduce the uncertainty from the CKM
matrix element, we propose a quantity, defined by the ratio
\begin{eqnarray}
R(B^{-}\to \tau \nu_{\tau}) = {BR(B^{-}\to \tau \nu_{\tau}) \over
BR(\bar B \to \pi^{+} e^{-} \bar\nu_{e})}\, . \label{eq:r1}
\end{eqnarray}
It is clear that the ratio of $R(B^{-}\to \tau \nu_{\tau})$ in Eq.
(\ref{eq:r1}) could directly reflect the charged Higgs effect in
$B^{-}\to \tau \nu_{\tau}$ as the charged contribution to $\bar B
\to \pi^{+} e^{-} \bar\nu_{e}$ is suppressed.
However, we introduce  new theoretical
uncertainty
arising from the transition form factor
$f^{\pi}_{+}(q^2)$ defined by Eq.~(\ref{eq:bpff}). To see the
influence of uncertainty on the ratio $R(B^{-}\to \tau \nu_{\tau})$,
we use two different QCD approaches of the light-front quark model (LFQM) \cite{LFQM} and light cone
sum rules (LCSRs) \cite{LCSR}  to
estimate the form factor. With $|V_{ub}|=3.67\times 10^{-3}$, we get that
the former predicts $BR(\bar B \to \pi^{+} e^{-}
\bar\nu_{e})=1.25\times 10^{-4}$ while the latter $BR(\bar B \to
\pi^{+} e^{-} \bar\nu_{e})=1.55\times 10^{-4}$, which are consistent
with the data of $(1.33\pm 0.22)\times 10^{-4}$ \cite{PDG04}. From
the results, we see that the error from the uncertainty of
$f^{\pi}_{+}(q^2)$ on the $R(B^{-}\to \tau \nu_{\tau})$ could be around
$20\%$ which is still less than the error of $40\%$ from $V_{ub}$.
To be more clear, in Fig.~\ref{fig:error}(b) we display the ratio
$R(B^{-}\to \tau \nu_{\tau})$ by LFQM (dot-dashed) and LCSRs
(dot-dot-dashed) in the SM, where the solid and dashed lines denote
the central value and errors of the current data $R(B^{-}\to \tau
\nu_{\tau})=1.02\pm 0.40$, respectively.
%

\begin{figure}[htbp]
\includegraphics*[width=4.5 in]{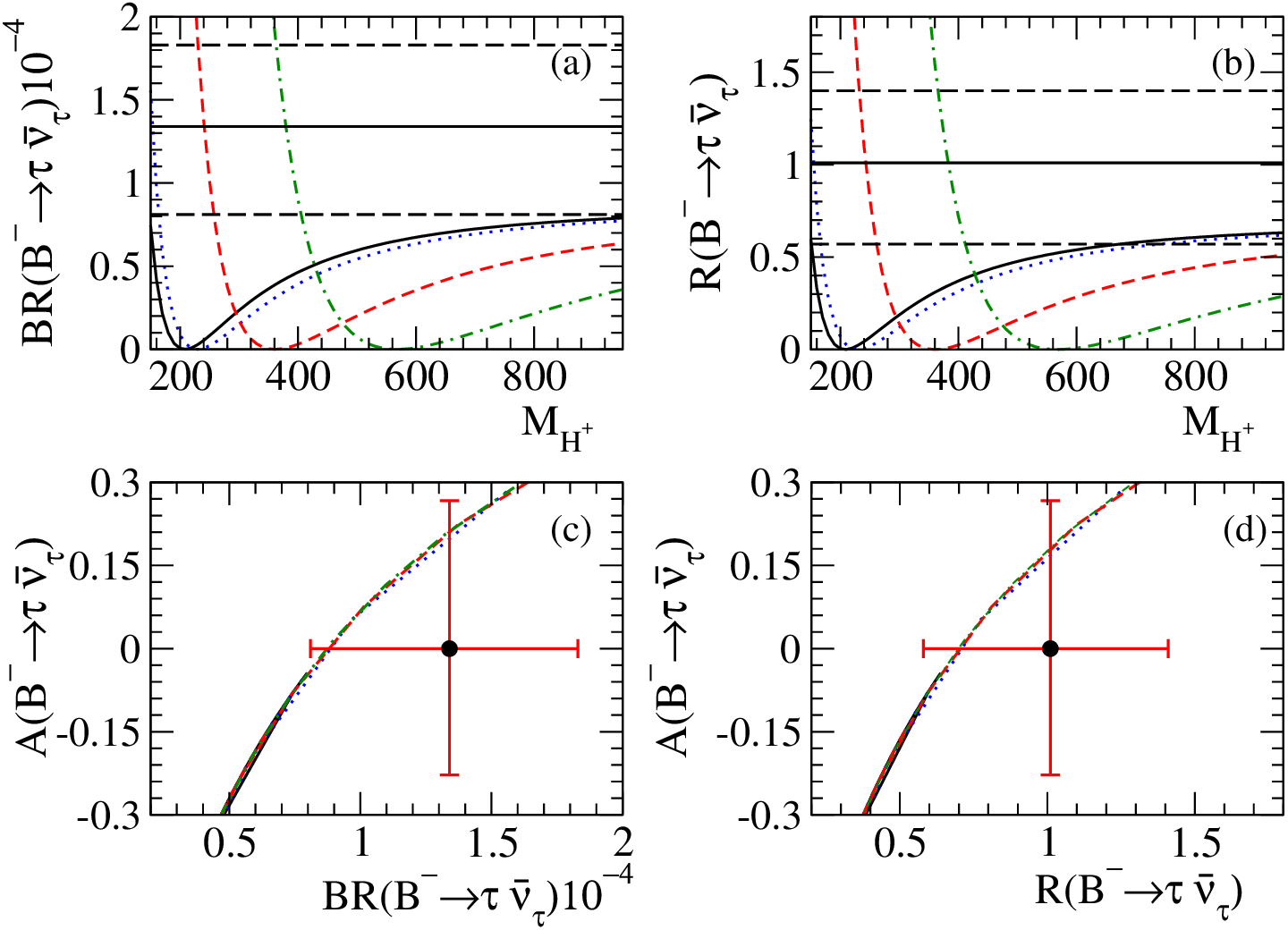}
\caption{(a) BR (in units of $10^{-4}$) and (b) $R$ for $B^{-}\to \tau
\bar\nu_{\tau}$ as a function of $M_{H^+}$
and  (c)[(d)] ${\cal A}(B^{-}\to
\tau \bar\nu_{\tau})$
with respect
to $BR(B^{-}\to \tau \bar\nu_{\tau})$ [$R(B^{-}\to \tau
\bar\nu_{\tau})$], where
the solid, dotted, dashed and dot-dashed lines
correspond to the possible sign combinations of $(e_{0},\, e_Y)$
denoted by $(+,+)$, $(+,-)$, $(-,+)$ and $(-,-)$, respectively, and
 the data
with errors are included.}
 \label{fig:btaunu_phys}
\end{figure}

\begin{figure}[htbp]
\includegraphics*[width=4.5 in]{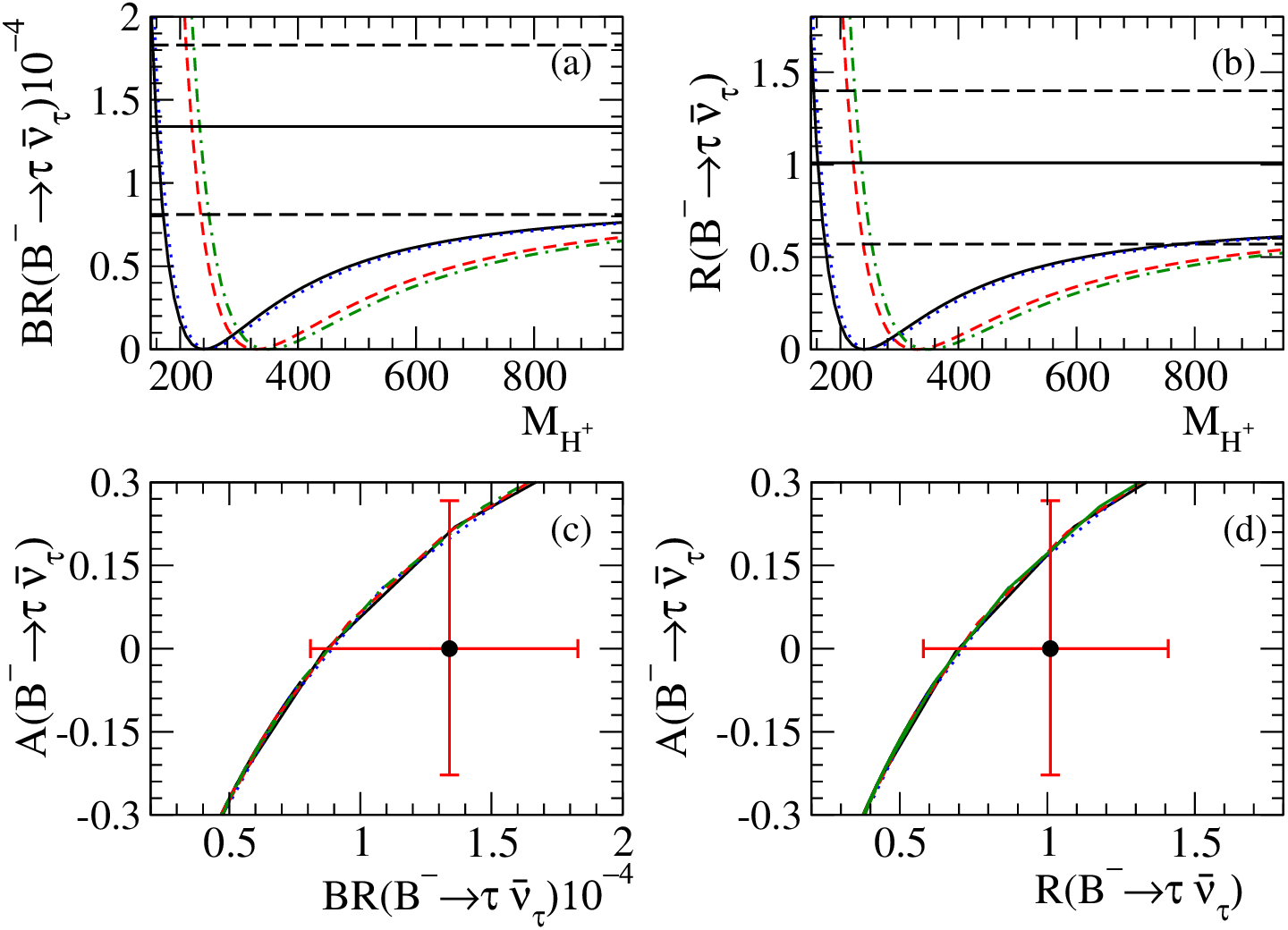}
\caption{ Legend is the same as Fig.~\ref{fig:btaunu_phys} but
$X=0.5$.}
 \label{fig:btaunu_phys2}
\end{figure}

In terms of Eq.~(\ref{eq:loopeff}), we now study the influence of
the charged Higgs. First of all, to understand how the charged Higgs
affects $B^{-}\to \tau \bar\nu_{\tau}$ directly, we display
$BR(B^{-}\to \tau \bar\nu_{\tau})$ as a function of the charged
Higgs mass in Fig.~\ref{fig:btaunu_phys}(a), where we have taken
$X=1$, $f_{B}=0.19$ GeV and $|V_{ub}|=3.67\times 10^{-3}$. Since
there is a two-fold ambiguity in sign for each $(e_{0},\, e_Y)$, the
solid, dotted, dashed and dot-dashed lines correspond to the
possible sign combinations denoted by $(+,+)$, $(+,-)$, $(-,+)$ and
$(-,-)$, respectively. From the figure, we see that the decay
$B^{-}\to \tau \bar\nu_{\tau}$ could exclude some parameter space.
To remove the uncertainty of $V_{ub}$, in
Fig.~\ref{fig:btaunu_phys}(b) we show the effects of the charged
Higgs on $R(B^{-}\to \tau \nu_{\tau})$.
In order to make the new physics effects more clearly, we define
another physical quantity as
\begin{eqnarray}
{\cal A}(B^{-}\to \tau \nu_{\tau})={R(B^{-}\to \tau \nu_{\tau})-
R^{SM}(B^{-}\to \tau \nu_{\tau})\over R(B^{-}\to \tau \nu_{\tau})+
R^{SM}(B^{-}\to \tau \nu_{\tau})}\, .\label{eq:a_phys}
\end{eqnarray}
Although the quantity $R(B^{-}\to \tau \nu_{\tau})$ still depends on
$f_{B}$ and $f^{\pi}_{+}(q^2)$, the new quantity ${\cal A}(B^{-}\to
\tau \nu_{\tau})$ reduces their
dependences. That
is, if a nonzero value of ${\cal A}(B^{-}\to \tau \nu_{\tau})$ is measured,
it shows the existence of new physics definitely. We present the
charged Higgs contributions   to ${\cal A}(B^{-}\to \tau
\nu_{\tau})$ with respect to $BR(B^{-}\to \tau \nu_{\tau})$ and
$R(B^{-}\to \tau \nu_{\tau})$ in Fig.~\ref{fig:btaunu_phys}(c) and
(d), respectively, where we also display the current bounds.
  Clearly, ${\cal A}(B^{-}\to \tau \nu_{\tau})\sim 10\%$ is
easy to reach by the charged Higgs effects in the MSSM. We note that
the new physical quantity ${\cal A}(B^{-}\to \tau \nu_{\tau})$ is
not sensitive to the signs in $e_{0}$ and $e_{Y}$. Similarly, we
show the results with $X=0.5$ in Fig.~\ref{fig:btaunu_phys2}.

\subsection{$\bar B \to (\pi^{+}, D^{+}) \ell \bar \nu_{\ell}$}

Besides the CKM matrix element, the main theoretical uncertainty for
$\bar B\to P \ell \bar\nu_{\ell}$  is from the $B\to P$
transition form factors. For numerical estimations, we employ the
results of the LFQM \cite{LFQM} in which the form factors as a
function of $q^2$ are parametrized by
\begin{eqnarray}
f^{P}(q^2)= {f^{P}(0) \over 1- a q^2/m^2_{B} +b(q^2/m^2_{B})^2}\, ,
\end{eqnarray}
and the fitting values of parameters $a$ and $b$ are
shown in Table~\ref{table:F_P}.
\begin{table}[hptb]
\caption{ The transition form factors for $B\to (\pi,\, D)$
calculated by  the LFQM \cite{LFQM}.}\label{table:F_P}
\begin{ruledtabular}
\begin{tabular}{cccccccc}
$f^{P}(q^2)$ & $f^{P}(0)$ &$a$& $b$ & $f^{P}(q^2)$ & $f^{P}(0)$ & $a$ & $b$ \\ \hline 
 $f^{\pi}_{+}(q^2)$ &  $0.25$ & $1.73$ & $0.95$ &  $f^{\pi}_{0}(q^2)$ &  $0.25$ & $0.84$ & $0.10$  \\ \hline
 $f^{D}_{+}(q^2)$ & $0.67$ & $1.25$ & $0.39$ & $f^{D}_{0}(q^2)$ & $0.67$ & $0.65$ & $0.00$
 \\
 \end{tabular}
\end{ruledtabular}
\end{table}
To check the contributions of the input form factors in the SM, we
present BRs for $\bar B\to (\pi^{+},\, D^{+}) \ell \bar\nu_{\ell}$
in Table~\ref{table:br_P}, where we have used $|V_{ub}|=3.67\times
10^{-3}$ and $|V_{cb}|=(41.3\pm 0.15)\times 10^{-3}$ \cite{PDG04}.
It is clear that for the light lepton production, the results are
consistent with the data. Since the new coupling of the charged
Higgs is associated with the lepton mass, it is easily to understand
that the effects of the charged Higgs will not significantly affect
the light leptonic decays. Hence, in our analysis, we will only
concentrate on the $\tau$ decay modes.
 \begin{table}[hptb]
\caption{BRs for $\bar B\to \pi^{+} \ell^{-} \bar \nu_{\ell}$ with
$|V_{ub}|=(3.67\pm 0.47)\times 10^{-3}$  and $\bar B\to D^{+}
\ell^{-} \bar \nu_{\ell}$ with $|V_{cb}|=(41.3\pm 1.5)\times
10^{-3}$ in the SM. }\label{table:br_P}
\begin{ruledtabular}
\begin{tabular}{cccccc}
Mode & $\bar B\to \pi^{+} \ell^{-} \nu_{\ell}$ & $\bar
B\to \pi^{+} \tau^{-} \nu_{\tau}$  & $\bar B\to D^{+}
\ell^{-} \nu_{\ell}$ &
$\bar B\to D^{+} \tau^{-} \nu_{\tau}$ \\ \hline 
  SM &  $(1.25\pm 0.23) 10^{-4}$ & $(0.85\pm 0.15)10^{-4}$ & $(2.29\pm 0.12)\%$  &  $(0.69\pm 0.04 )\%$\\ \hline
Experiment \cite{PDG04} &  $(1.33\pm 0.22) 10^{-4} $ & $$ & $(2.12\pm 0.20)\%$ &   \\
 \end{tabular}
\end{ruledtabular}
\end{table}

\begin{figure}[htbp]
\includegraphics*[width=4.5 in]{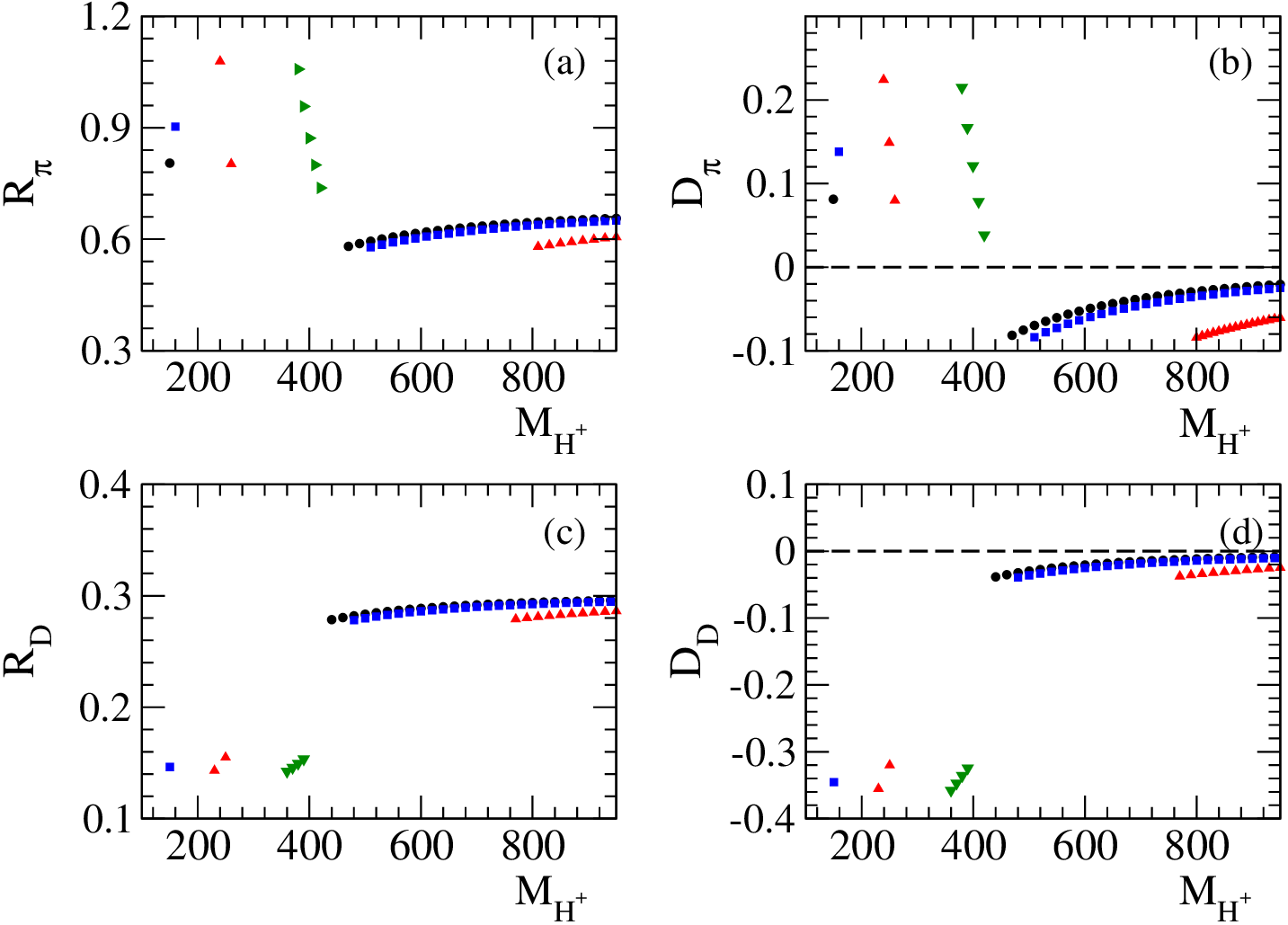}
\caption{(a)[(c)] denotes the $R_{\pi}[R_{D}]$ and (b)[(d)] displays
$D_{\pi[D]}$ with respect to $M_{H^+}$ for $X=1$, where the circle,
square, triangle-up and triangle down represent the sign
combinations of $(e_0,\, e_Y)$ such as $(+,+)$, $(+,-)$, $(-,+)$ and
$(-,-)$.}
 \label{fig:brbplnu}
\end{figure}

\begin{figure}[htbp]
\includegraphics*[width=4.5 in]{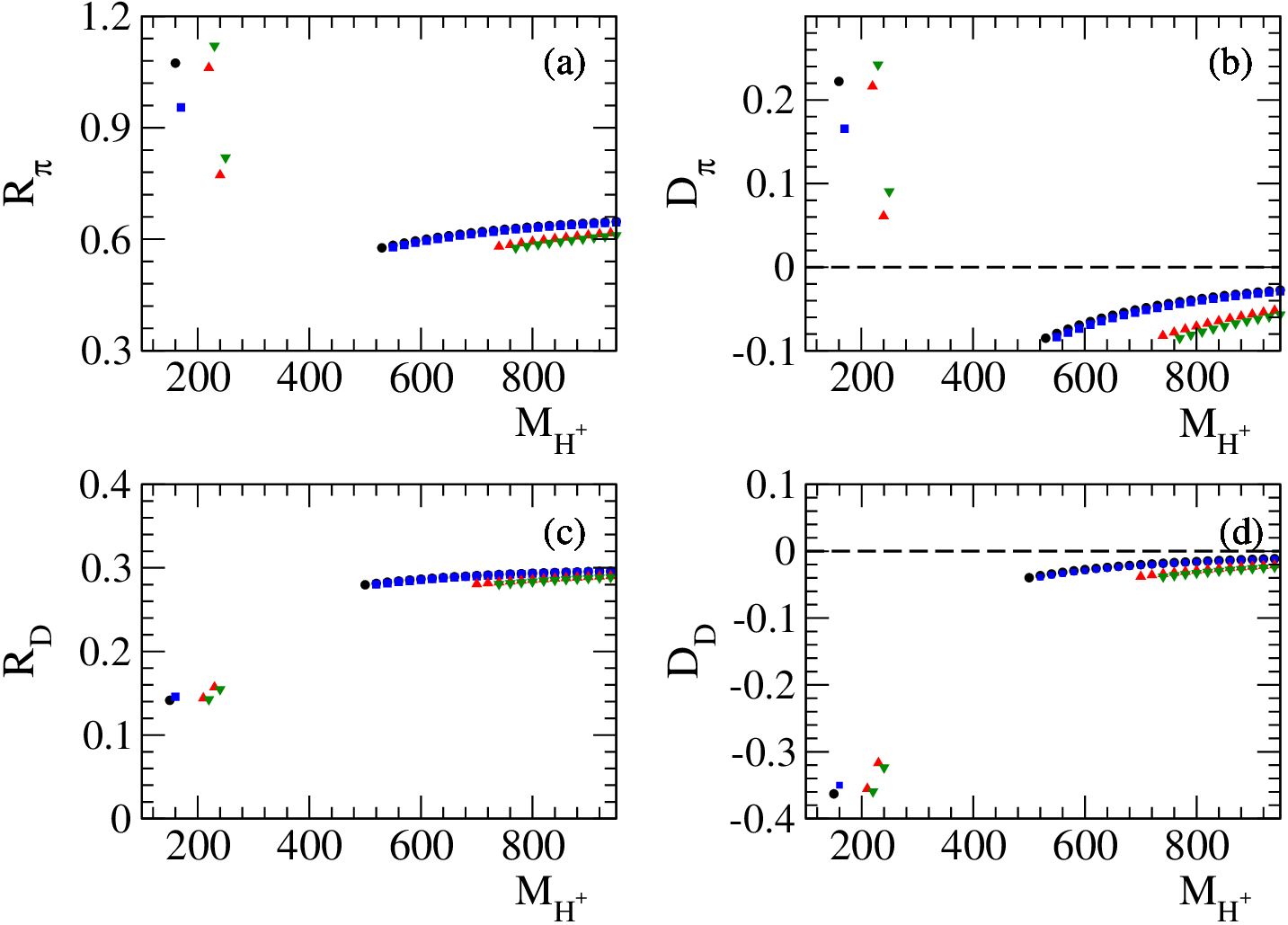}
\caption{ Legend is the same as Fig.~\ref{fig:brbplnu} but $X=0.5$.}
 \label{fig:brbplnu_x05}
\end{figure}

Since $\bar B\to (\pi^{+},\, D^{+}) \tau \bar \nu_{\tau}$ have not
been observed yet, we take  $R(B^{-}\to \tau \bar
\nu_{\tau})=1.02\pm 0.40$ as a constraint. To reduce the theoretical
uncertainty from the CKM matrix elements, we consider the ratio
\begin{eqnarray}
R_{P}={BR(\bar B\to P \tau \bar \nu_{\tau})\over BR(\bar
B\to P \ell \bar\nu_{\ell})}
\end{eqnarray}
instead of $BR(\bar B\to P \tau \bar \nu_{\tau})$.
In addition,
to illustrate new physics clearly,
we also define
\begin{eqnarray} {\cal D}_{P} = {
R_{P}-R^{SM}_{P} \over R_{P}+R^{SM}_{P}}\,. \label{eq:dev2}
\end{eqnarray}
If a non-zero value of ${\cal D}_{P}$ is observed, it must indicate the
existence of new physics. Hence, according to Eq.~(\ref{eq:diff_P}),
$R_{P}$ and ${\cal D}_{P}$ for $P=(\pi^{+},\, D^{+})$ with $X=1$
are displayed in Fig.~\ref{fig:brbplnu}, where the circle, square,
triangle-up and triangle-down symbols correspond to the possible
signs for $e_{0}$ and $e_{Y}$, expressed by $(+,+)$, $(+,-)$,
$(-,+)$ and $(-,-)$, respectively. Similar analysis with $X=0.5$ is
presented in Fig.~\ref{fig:brbplnu_x05}. By the figures, we see that
the input $R(B^{-}\to \tau \bar\nu_{\tau})$ has given a strict
constraint on the
signs of $e_{0}$ and $e_{Y}$ and
the parameters of $X=|\mu|/M_{S}$ and $M_{H^+}$. Even so, we still can have $O(10\%)$
deviation in ${\cal D}_{P}$ when $M_{H^+}$ is less than $400$ GeV.
%
\begin{figure}[htbp]
\includegraphics*[width=4.5 in]{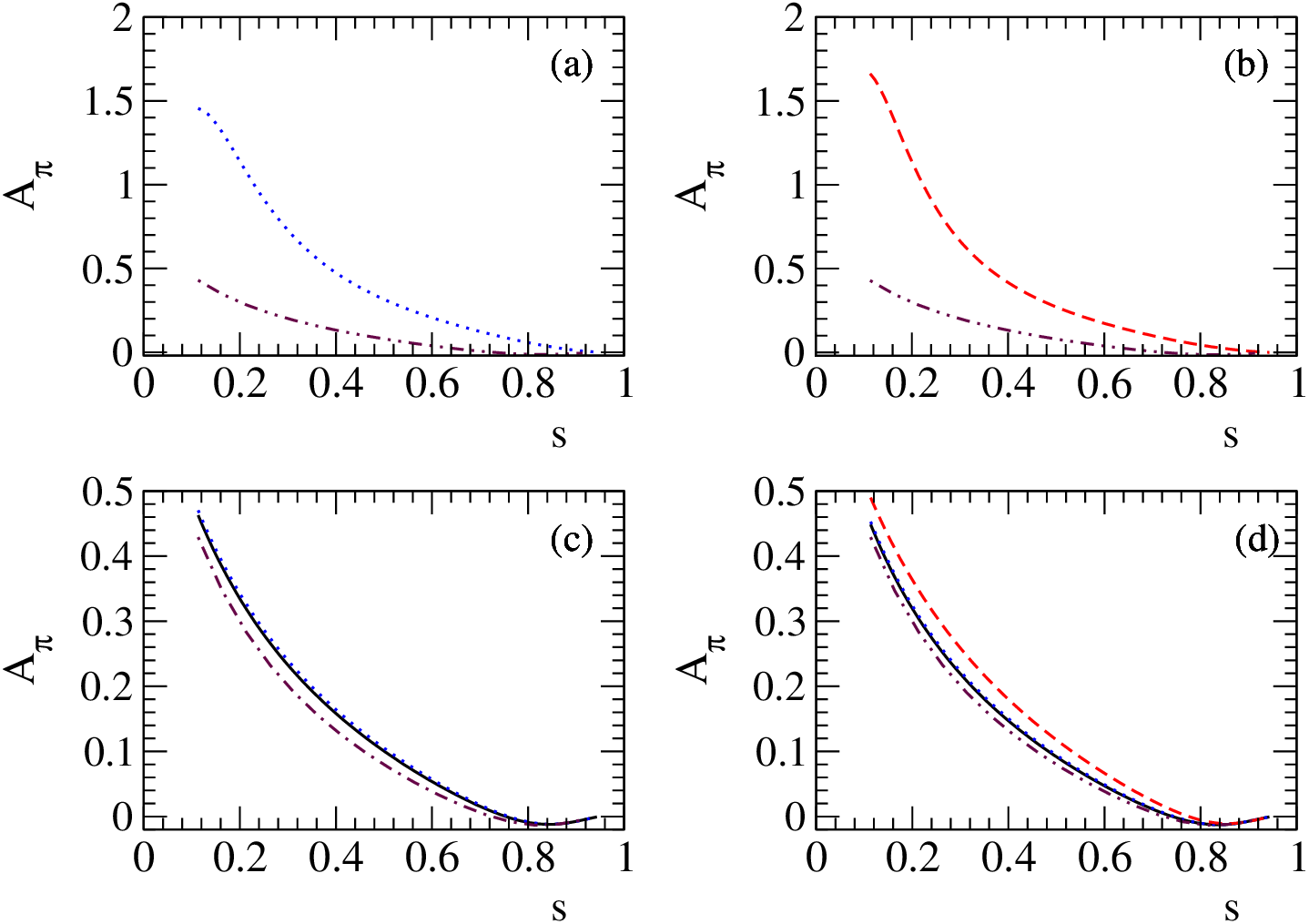}
\caption{
Angular asymmetries
for $\bar B \to \pi^{+} \tau \bar\nu_{\tau}$  with
$M_{H^+}=$ (a) 160 GeV, (b) 230 GeV, (c) 650 GeV and (d) 850 GeV,
where the solid, dotted, dashed, dash-dotted lines correspond
to the sign combinations of $(e_{0},\, e_{Y})$,
 expressed by $(+,+)$, $(+,-)$, $(-,+)$, $(-,-)$,
 respectively, and the dash-dotted-dotted lines represent the
 SM results.
}
 \label{fig:asypi_x1}
\end{figure}
\begin{figure}[htbp]
\includegraphics*[width=4.5 in]{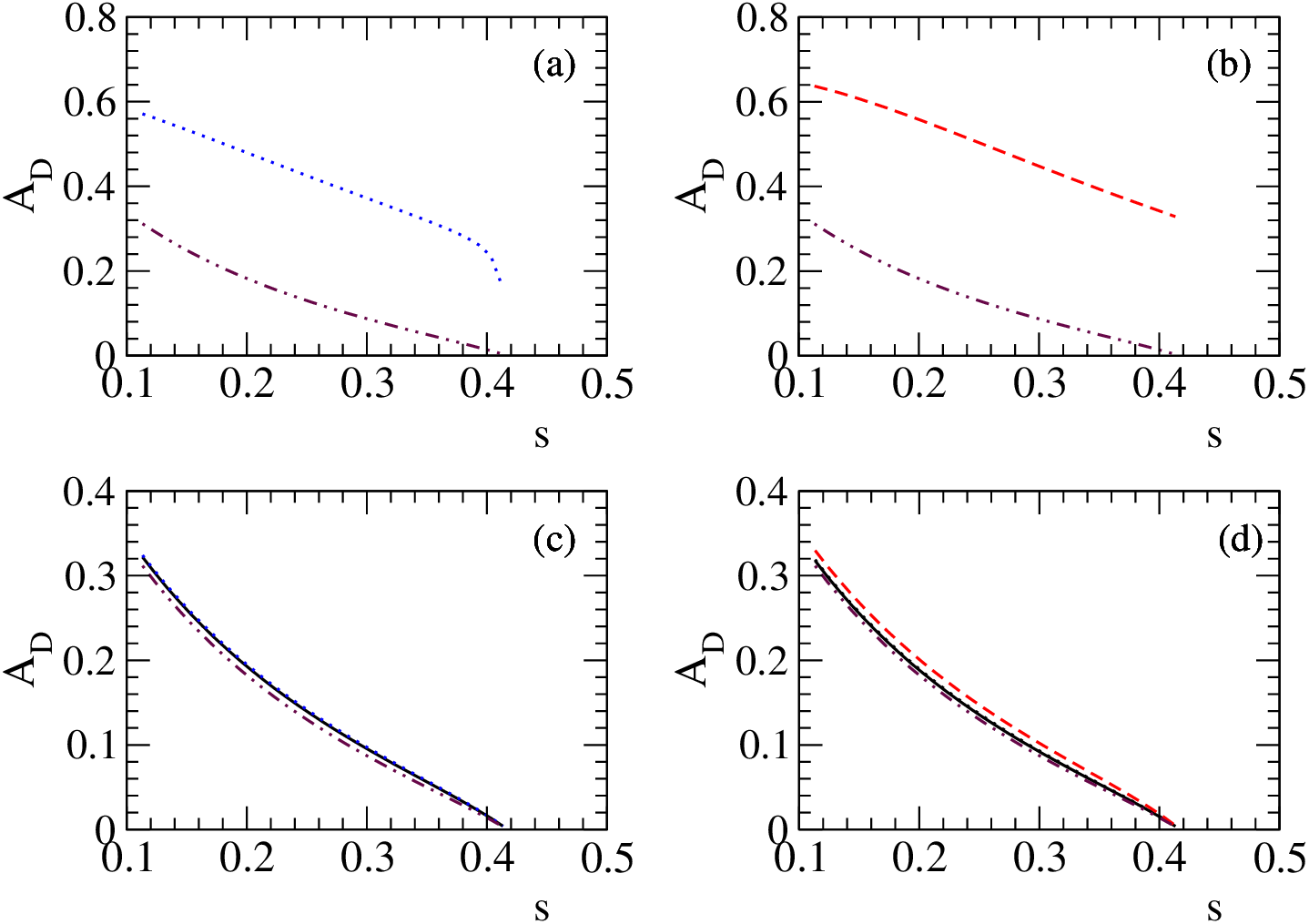}
\caption{Legend is the same as Fig.~\ref{fig:asypi_x1} but for $\bar
B\to D^{+} \tau \bar\nu_{\tau}$.}
 \label{fig:asyD_x1}
\end{figure}

Note that apart from the BR related quantities,
 the angular
 distribution asymmetry defined in Eq.~(\ref{asy_P}) could also
 be used to
examine the effects beyond the SM \cite{CG_PRD71}. We display the
contributions of the charged Higgs with $X=1$ to ${\cal A}_{P}$ in
Fig.~\ref{fig:asypi_x1} (\ref{fig:asyD_x1}) for $\bar B\to
\pi^{+} \tau \bar\nu_{\tau}$ ($\bar B\to D^{+} \tau
\bar\nu_{\tau}$). In the figures,  (a), (b), (c) and (d)
correspond to $M_{H^+}=160,\, 230,\, 650$ and  $850$ GeV,
 the solid, dotted, dashed and dot-dashed denote the
sign combinations of $(e_0,\, e_Y)=(+,+)$, $(+,-)$,
$(-,+)$ and $(-,-)$, and the dash-dotted-dotted line stands for the
SM result, respectively.
 From the figures, it is clear that the charged Higgs
contributions in the light $M_{H^+}$ region could significantly
affect the angular asymmetries. We note that due to the
constraint $R(B^{-}\to \tau \nu_{\tau})=1.02\pm 0.40$, some
sign combinations have been excluded with the same $M_{H^+}$.

\subsection{ $\bar B \to (\rho^{+}, D^{*+}) \ell \bar \nu_{\ell}$}

The form factors in $B\to (\rho,\, D^*)$ are parametrized by
\begin{eqnarray}
f^{V}(q^2)= {f^{V}(0) \over 1- a q^2/m^2_{B} +b(q^2/m^2_{B})^2}
\end{eqnarray}
with $a$ and $b$ given in
Table~\ref{table:F_V}.
 \begin{table}[hptb]
\caption{The transition form factors for $B\to (\rho,\, D^{*})$
calculated by the LFQM  \cite{LFQM}.}\label{table:F_V}
\begin{ruledtabular}
\begin{tabular}{cccccccc}
$f^{V}(q^2)$ & $f^{V}(0)$ &$a$& $b$ & $f^{V}(q^2)$ & $f^{V}(0)$ & $a$ & $b$ \\ \hline 
$V^{\rho}(q^2)$ &  $0.27$ & $1.84$ & $1.28$ &  $A^{\rho}_{0}(q^2)$
&  $0.28$ & $1.73$ & $1.20$  \\ \hline $A^{\rho}_{1}(q^2)$ &
$0.22$ & $0.95$ & $0.21$ & $A^{\rho}_{2}(q^2)$ & $0.20$ & $1.65$&
$1.05$ \\ \hline
$V^{D^*}(q^2)$ &  $0.75$ & $1.29$ & $0.45$ & $A^{D^*}_{0}(q^2)$ &
$0.64$ & $1.30$ & $0.31$  \\ \hline $A^{D^*}_{1}(q^2)$ & $0.63$ &
$0.65$ & $0.02$ & $A^{D^*}_{2}(q^2)$ & $0.61$ & $1.14$& $0.52$
 \\
 \end{tabular}
\end{ruledtabular}
\end{table}
Based on these form factors and  Eq.~(\ref{eq:diff_V}),
the  BRs in the SM are shown in
Table~\ref{table:br_V}.
 \begin{table}[hptb]
\caption{BRs for $\bar B\to \rho^{+} \ell^{-} \bar \nu_{\ell}$ with
$|V_{ub}|=(3.67\pm 0.47)\times 10^{-3}$  and $\bar B\to D^{*+}
\ell^{-} \bar \nu_{\ell}$ with $|V_{cb}|=(41.3\pm 1.5)\times
10^{-3}$ in the SM. }\label{table:br_V}
\begin{ruledtabular}
\begin{tabular}{cccccc}
Mode & $\bar B\to \rho^{+} \ell^{-} \nu_{\ell}$ & $\bar
B\to \rho^{+} \tau^{-} \nu_{\tau}$  & $\bar B\to D^{*+}
\ell^{-} \nu_{\ell}$ &
$\bar B\to D^{*+} \tau^{-} \nu_{\tau}$ \\ \hline 
  SM &  $(3.18\pm 0.56) 10^{-4}$ & $(1.73\pm 0.31)10^{-4}$ & $(5.60 \pm 0.29)\%$  &  $(1.41\pm 0.07)\%$\\ \hline
Exp \cite{PDG04} &  $(2.6\pm 0.7) 10^{-4} $ & $$ & $(5.34\pm 0.20)\%$ &   \\
 \end{tabular}
\end{ruledtabular}
\end{table}
It is clear that for the light lepton production, the BRs are
consistent with the current experimental data. By using the same
form factors to the processes asscoated with the $\tau$ production, if
any significant deviation from the predictions of the SM is found,
it should indicate new physics.

\begin{figure}[htbp]
\includegraphics*[width=4.5 in]{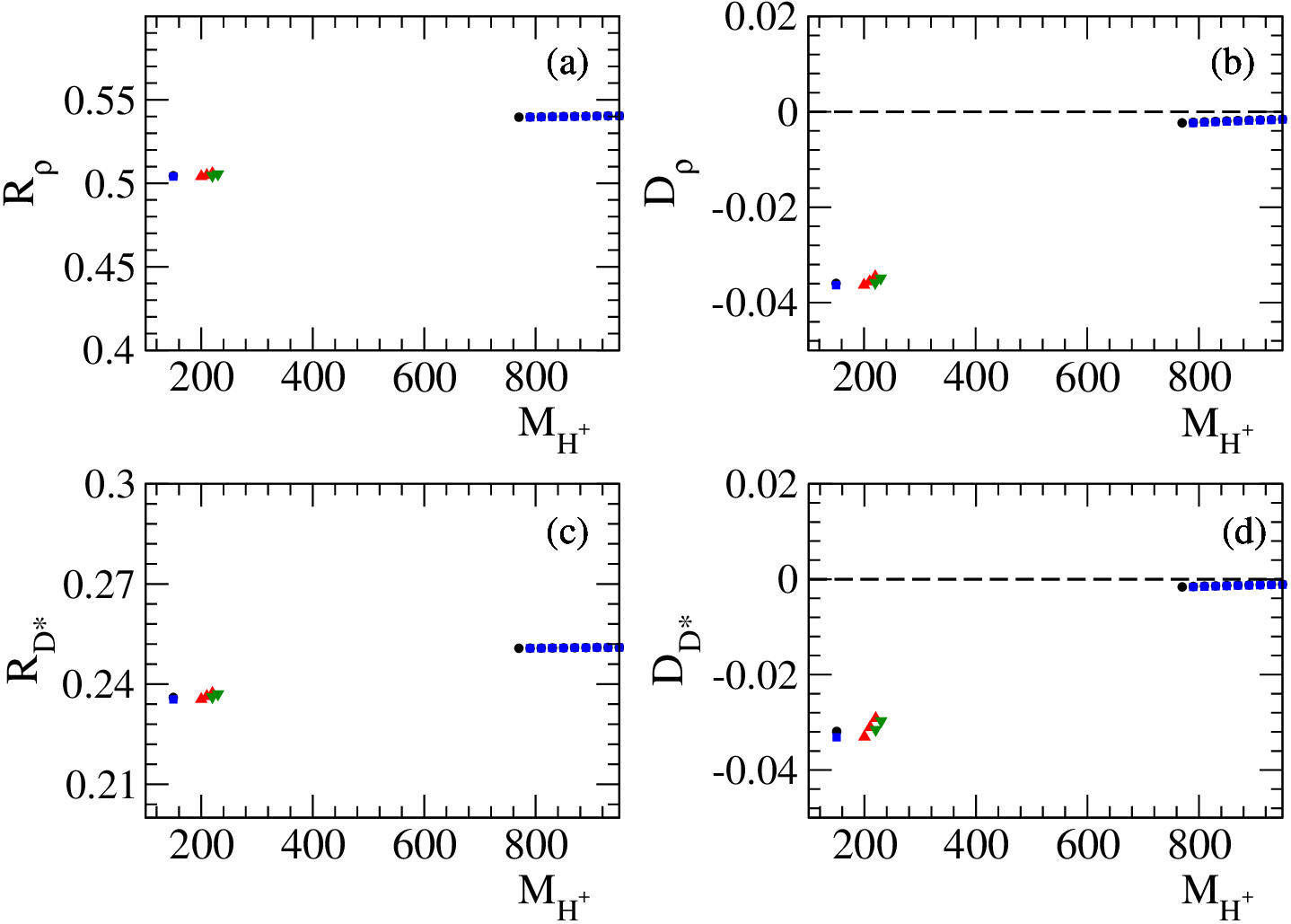}
\caption{(a)[(c)] denotes  $R_{\rho}[R_{D^*}]$ and (b)[(d)]
displays $D_{\rho[D^*]}$ as functions of $M_{H^+}$ for $X=0.5$
with $R(B^-\to \tau \bar\nu_{\tau})=1.02\pm
0.40$.
Legend is the same as Fig.~\ref{fig:brbplnu}.
}
 \label{fig:brbrholnu}
\end{figure}

\begin{figure}[htbp]
\includegraphics*[width=4.5 in]{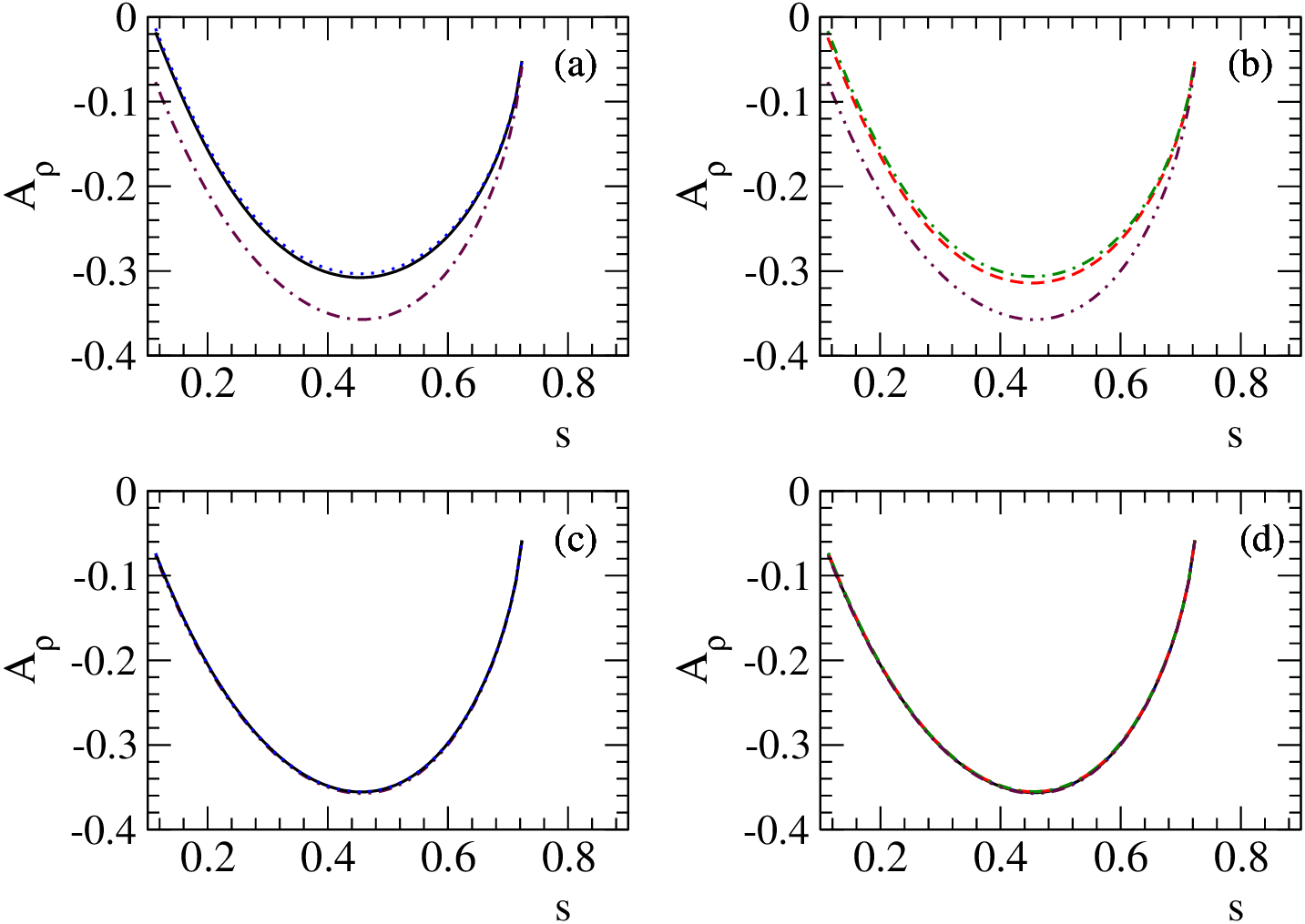}
\caption{Legend is the same as Fig.~\ref{fig:asypi_x1} but for $\bar B\to \rho^{+} \tau \bar\nu_{\tau}$.}
 \label{fig:asyrho}
\end{figure}

\begin{figure}[htbp]
\includegraphics*[width=4.5 in]{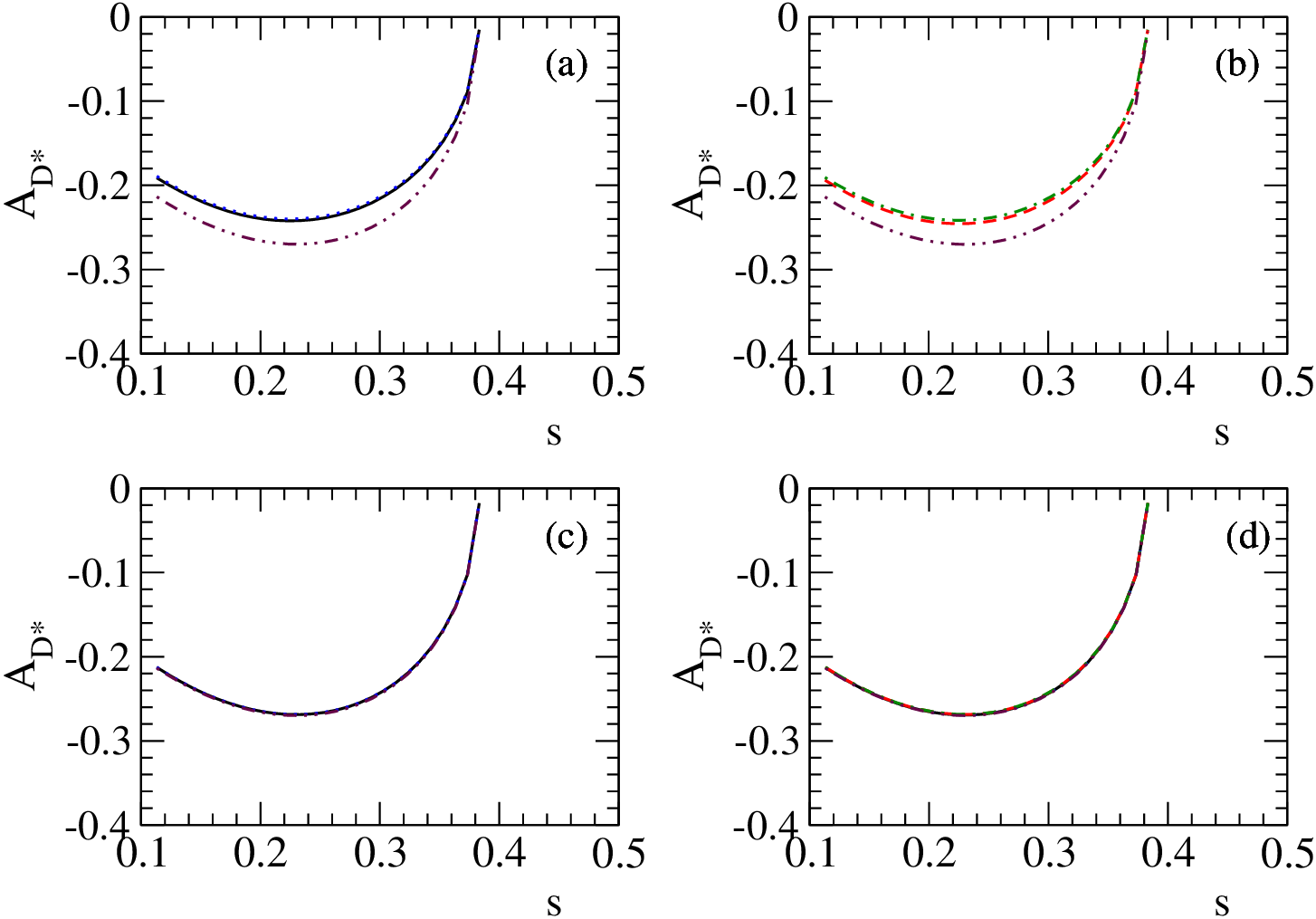}
\caption{Legend is the same as Fig.~\ref{fig:asypi_x1} but for $\bar B\to D^{*+} \tau \bar\nu_{\tau}$.}
 \label{fig:asyDs}
\end{figure}

Similar to the decays $\bar B\to (\pi^{+},\, D^{+}) \ell
\bar\nu_{\ell}$,
we define
\begin{eqnarray}
R_{V}={BR(\bar B\to V \tau \bar \nu_{\tau})\over BR(\bar
B\to V \ell \bar\nu_{\ell})}\ \ {\text and}\ \ {\cal
D}_{V}={R_{V}-R^{SM}_{V}\over R_{V}+R^{SM}_{V}}\, .
\end{eqnarray}
The  contributions of the charged Higgs are presented
in Fig.~\ref{fig:brbrholnu} with $X=0.5$. To constrain the free
parameters, we have taken $R(B^-\to \tau \bar\nu_{\tau})=1.02\pm
0.40$. From the results, we see that the  charged Higgs
contributions to ${\cal D}_{V}$ are only at few
percent.
In addition, we also display the angular asymmetries for
$\bar B\to (\rho^{+},\, D^{+*}) \tau \bar \nu_{\tau}$ in
Figs.~\ref{fig:asyrho} and \ref{fig:asyDs}.
We find
that the influence of the light charged Higgs on ${\cal A}_{\rho}$ is
larger than that on ${\cal A}_{D^{*}}$. However, the contributions
 from the heavy charged Higgs are the same as the predictions in the SM.

Finally, we make some comparisons in $\bar B\to P \tau
\bar\nu_{\tau}$ and $\bar B\to V \tau \bar\nu_{\tau}$.
For $\bar B\to P \ell
\bar\nu_{\ell}$, according to
Eq.~(\ref{eq:diff_P}), one finds that the dominant effects for the
BRs, which do not vanish in the limit of $m_{\ell}=0$, are
$\propto f^{P2}(q^2) \hat P^{2}_{P}$. Although the terms directly
related to the lepton mass  in the form of
$\hml^{2}f^{P2}_{+}(q^2)$, for the $\tau$ modes, the mass effects
 could have $O(10\%)$ in order of magnitude. Since the
new charged Higgs contributions appear in the terms associated with
$\hml^{2}f^{P2}_{0}(q^2)$, it is expected that in average the
influence of the charged Higgs could be as large as $O(10\%)$, which
is consistent with the results shown in the
Fig.~\ref{fig:brbplnu}. Furthermore, since the lepton angular
asymmetry is associated with $\hml^2 f^{P}_{+,\,0}(q^2)$, we can
understand that ${\cal A}_{P}$, shown in the
Figs.~\ref{fig:asypi_x1} and \ref{fig:asyD_x1}, could be
significantly affected by the charged Higgs couplings. However, the
situation is different in the decays $\bar B\to V \tau
\bar\nu_{\tau}$. Since the vector meson carries spin degrees of freedom,
besides longitudinal parts which are similar to $\bar B\to P \tau
\bar\nu_{\tau}$, there also exist transverse contributions.
Therefore, the effects $\propto \hml^{2}f^{P2}_{0}(q^2)$ become
relatively small. This is the reason why the results in
Figs.~\ref{fig:brbrholnu}, \ref{fig:asyrho} and \ref{fig:asyDs} are
not sensitive to the charged Higgs effects.
We conclude that the charged Higgs contributions on
$\bar B\to V \ell \bar \nu_{\ell}$ are much less than those on $\bar
B\to P \ell \bar\nu_{\ell}$.

\section{Conclusion} \label{sec:conclusion}
Motivated by the recent measurement on the decay branching ratio of
$B^{-} \to \tau \bar\nu_{\tau}$, we have studied the exclusive
semileptonic decays of $\bar B\to  (\pi,D,\rho,D^{*})^{+} \ell
\bar\nu_{\ell}$ in the MSSM. In particular, we have examined the
charged Higgs effects from nonholomorphic terms at the large value
of $\tan\beta$. To extract new physics contributions, we have
defined several physical quantities to reduce uncertainties from the
QCD as well as the CKM elements. Explicitly, for the allowed region
of the charged Higgs mass, with the constraints from $BR(B^{-} \to
\tau \bar\nu_{\tau})$ we have shown that ${\cal A}(B^{-} \to \tau
\bar\nu_{\tau})$ and ${\cal D}_{\pi,D}\sim 10\%$ are still allowed,
whereas $D_{\rho,D^{*}}$ are small. Moreover, we  have demonstrated
that the angular asymmetries of ${\cal A}_{\pi,D}$ could be
significantly enhanced in the light $M_{H^+}$ region, whereas those
of ${\cal A}_{\rho,D^{*}}$ are insensitive to the charged Higgs
contributions. It is clear that if one of the above physical
quantities
is observed, it is a signature of new physics, such as the charged Higgs.\\
\\
{\bf Acknowledgments}\\

The authors would like to thank  Dr. Kai-Feng Chen for useful
discussions. This work is supported in part by the National Science
Council of R.O.C. under Grant \#s:NSC-95-2112-M-006-013-MY2,
NSC-94-2112-M-007-(004,005) and NSC-95-2112-M-007-059-MY3.

\end{document}